\documentclass[aps,twocolumn,nofootinbib]{revtex4}
\usepackage[utf8x]{inputenc}
\usepackage{epsfig}
\usepackage{amsfonts}
\usepackage{amssymb}
\usepackage{amsmath}
\newcommand{\beq}{\begin{equation}}
\newcommand{\eeq}{\end{equation}}
\newcommand{\kB}{k_{\mbox{\tiny B}}}
\usepackage{physics}

\begin{document}

\title{Entropy production  and heat transport in harmonic chains
  under time dependent periodic drivings}

\author{Bruno A. N. Akasaki, Mário J. de Oliveira and C. E. Fiore}
\affiliation{Universidade de São Paulo,
Instituto de Física,
Rua do Matão, 1371, 05508-090
São Paulo, SP, Brasil}
\date{\today}
\begin{abstract}
  Using  stochastic thermodynamics, the properties of interacting
  linear chains subject to  periodic drivings are investigated.
  The systems are described by Fokker-Planck-Kramers equation and exact (explicit) solutions
  are   obtained for periodic drivings as  functions of the modulation frequency
  and strength constants.
The limit of long chains is analyzed by means
of a protocol in which the intermediate temperatures
are self-consistently chosen and  the entropy production is decomposed as a sum of
two individual contributions, one coming from real baths (placed at extremities
of lattice) and other from self-consistent baths. The thermal
reservoirs  lead to a heat flux according to Fourier's law.
Explicit expressions for short
chains are derived, whose entropy production is written down as a bilinear
function of thermodynamic forces and the associated fluxes, from which
Onsager coefficients have been evaluated.
A comparison between distinct periodic drivings is also performed.
\end{abstract}

\maketitle

\section{Introduction}

The description  of  thermodynamic
quantities at the microscopic level gives  rise to the
 stochastic thermodynamics
\cite{prigo,groot,mariobook}, in which
fluctuations in the thermodynamic fluxes become important.
This theory  not only   allows to reproduce the fundamental
concepts of thermodynamics of equilibrium systems but
can also be extended for the more general case of nonequilibrium ones.
In particular, it shows that stochastic fluxes satisfy general
relations such as the Jarzynski equality \cite{j1,j2} or/and
it predicts the existence of general bounds among thermodynamic fluxes \cite{kau,barau}.

In the last years, the concept of
entropy production has played a fundamental role in nonequilibrium
statistical physics not only for
typifying the irreversibility \cite{seifert12,tome2015,broeck15}, but also
for  tackling general considerations about efficiency of heat engines \cite{karel2016prl}, the analysis of (irreversible)  phase transition  portraits \cite{noa,esposito1,esposito2}, thermodynamic uncertainties relations \cite{kau,barau}
and others.
A fundamental relation for the entropy production comes from  simple
entropic arguments in which the system is coupled to a thermal
reservoir. Its time entropy variation $dS/dt$ is the difference
of two terms
\begin{equation}
  \frac{dS}{dt} = \Pi(t) - \Phi(t),
  \label{eqs1}
\end{equation}
where $\Pi(t)$ is the entropy production rate  and $\Phi(t)$ is the entropy flux rate from/to the system to/from the environment. Since the environment works
a  subsystem in equilibrium, $\Pi(t)$ corresponds to the entropy produced
inside the system. Eq. (\ref{eqs1}) implies that   all
entropy spontaneously
produced (by the system) has to be  delivered to the environment in the steady state regime.
When the system is in thermal equilibrium,  it follows that $\Pi_s=\Phi_s=0$,
whereas $\Pi_s=\Phi_s>0$ out of the equilibrium regime. Thereby entropy production discerns equilibrium and nonequilibrium systems, since in the latter case
it is continuously produced.
In such
case, the steady entropy production rate can   alternatively be evaluated through the
calculation of the steady entropy flux $\Phi_s$.

The thermodynamic properties of   Markovian systems
have been extensively studied in the last years, including those
described by master  \cite{broeck20101,tome2015,broeck15,seifert12,ray} and  Fokker-Planck
equations \cite{tome2006,tome2010,broeck2010,rosas16,fiore19,karelc}.
 A special recent attention has been devoted to periodically
 driven systems \cite{seifert15,karel2016,fiorek,cleuren19}.
In part because
  their thermodynamic properties
  can be experimentally accessible \cite{seifert15,hohne2003,kraftmakher2004,filippov1966,sullivan1968,gobrecht1971,birge1985,gill1993,schawe1995,jeong1997,schawe1997,hohne1997,simon1997,jones1997,baur1998,claudy2000,garden2007a,garden2007c,garden2008}. In addition, some of their remarkable features,
  such as a general description in the linear regime (Onsager
  coefficients and general reciprocal relations can be achieved), the existence of uncertainties constraints leading
  to existence of bounds among macroscopic averages and other features
  have been  extensively
  studied.  However, the more general case of interacting particles
subject to time periodic drivings has not been studied thoroughly.

In this paper, we fill this gap by investigating the thermodynamic
properties of interacting chains of Brownian particles subject
to (time dependent) periodic forces and temperature drivings.
 Exact expressions for the
thermodynamic quantities, including the
dissipated heat,  entropy production, heat flux and others are obtained.
The  limits of short and long particle chains are thoroughly investigated.
For the latter case,
intermediate  temperatures are self-consistently chosen through a
protocol
taking into account an inner entropy production
source. This is meaningfully different from the original
approach by Bosterli et al. \cite{boster},
in which no heat flux is exchanged among the particles and self-consistent baths. Thus, our approach provides us not only to analyze the
role of two distinct sources
of dissipation,
but also establishing which contribution
dominates in the limits of short and long chains.
Two main findings can be drawn:
For the  case of two interacting particles,
the entropy production is derived as a bilinear function of fluxes and forces
for both drivings in  forces and temperatures, whose associated Onsager
coefficients depend on the interaction parameters and frequency driving. The entropy
production of long particle
chains can be splitted in two terms: one coming from the thermal
reservoirs and the other from the self-consistent ones.
Other remarkable features, such as the effect of a phase difference (a lag) between external forces are  investigated and
the condition for maximum/minimum entropy production    is found to depend only on the temperature
reservoirs and   frequency driving, irrespective the interaction
strength between particles.

This paper is organized as follows: In Sec. II we describe the theoretical
background and exact solution for time dependent drivings is evaluated
in Sec. III. In Sec. IV and V, the two and several particles cases
are analyzed, respectively. Conclusions are discussed in Sec. VI.

\section{Fokker-Planck-Kramers equation}

We consider a set of $N$ interacting particles  with equal
masses $m$, in which the $i$-th particle evolves in time according to the following set of coupled
Langevin equations
\begin{equation}
    \frac{d v_{i}}{dt} = f^{*}_{i} - \gamma v_{i} + \mathcal{F}_{i}(t),
  \label{two_baths_mov1}
\end{equation}
and
\begin{equation}
   \frac{d x_{i}}{dt} =  v_{i},
  \label{two_baths_mov2}
\end{equation}
 with $x_{i}$  denoting its position
 with velocity $v_{i} =d x_{i}/dt$, respectively, whereas
 $\gamma$ is the dissipation constant. Here $f^{*}_i$ stands for the
force acting to the $i$-th particle, which is assumed  to be
decomposed as the sum of a time dependent term $f_i^{ext}(t)$
plus a term
$f_{i}$ depending only on the positions. Thus,
 $f_{i}$ can be written as
the derivative of the potential energy $V$, $f_i=-\partial V/\partial x_i$.
The stochastic force $\mathcal{F}_{i}(t)$ accounts for the interaction between
particle $i$ and  the environment and satisfies the properties
\begin{equation}
 \langle \mathcal{F}_{i}(t)\rangle=0,
  \label{two_baths_ruido1}
\end{equation}
and
\begin{equation}
  \langle\mathcal{F}_{i}(t)\mathcal{F}_{j}(t^{'}) \rangle= 2 \gamma T_{i} \delta_{ij} \delta (t - t^{'}),
  \label{two_baths_ruido2}
\end{equation}
respectively,  where $T_i>0$ is distinct for each particle.
Let $P(x,v,t)\equiv P(x_{1},.., x_{N}, v_{1},..., v_{N}, t)$
be the joint probability distribution
 at time $t$, where $x$
and $v$
denote the collection of particle positions $x_i$ and velocities $v_i$,
respectively. Its time evolution
is described by the Fokker-Planck-Kramers (FPK) equation
\cite{mariobook,tome2010,tome2015}
\beq
\frac{\partial P}{\partial t} = - \sum_i\left(
v_i\frac{\partial P}{\partial x_i}
+ [f_i+f_i^{ext}(t)] \frac{\partial P}{\partial v_i}
+ \frac{\partial J_i}{\partial v_i}\right),
\label{3}
\eeq
where
\beq
J_i = - \gamma v_i P - \frac{\gamma \kB T_i}{m}
\frac{\partial P}{\partial v_i}.
\label{3a}
\eeq

If the temperatures of all particles $T_i$ are the same
and the external forces are null, the probability distribution
approaches  for large times the Gibbs equilibrium distribution,
\beq
P^{e}(x,v) = \frac{1}Z e^{-E/\kB T},
\label{9}
\eeq
where $E=mv^2/2 + V$ is the energy of the system.
This result shows that the FPK Eq. (\ref{3}) indeed
describes the contact of a system with a heat reservoir at a
temperature $T$. On the other hand, this will not be the
case of the system  in contact with distinct reservoirs and/or
when it is subject to time oscillating forces or temperatures.
In such case, the system dissipates heat and continuously
produce entropy.

From the FK equation, the time variation of the energy $U=\langle E\rangle$
reads
\beq
\frac{dU}{dt} = - \sum_{i=1}^{N}(\Phi^{(i)}_{\rm q}+ \Phi^{(i)}_{\rm w}),
\label{10}
\eeq
where the heat flux $\Phi^{(i)}_{\rm q}$  from the
system to the environment (thermal bath) is expressed as \cite{tome2010,tome2015}
\beq
\Phi^{(i)}_{\rm q} = \gamma( m \langle v_i^2\rangle -  \kB T_i),
\label{11}
\eeq
whose  first and second terms can be  understood as the
heating power and the power of heat losses, respectively.
The term $\Phi^{(i)}_{\rm w}$ can be interpreted as the work per unity of time
given by
\beq
\Phi^{(i)}_{\rm w} = -m \langle v_i\rangle f_i^{ext}(t).
\label{112}
\eeq
In the absence of external forces all heat flux comes from/goes to
the thermal bath.

The entropy $S$ of the system is determined from the
Gibbs expression
\beq
S = - \kB \int P \ln P dxdv.
\label{17}
\eeq
From the FPK equation, one finds that
its time derivative has the form of Eq. (\ref{eqs1}),
where the first is identified as the rate of entropy production
  given by \cite{tome2010,tome2015}
\beq
\Pi = \frac{m\kB}{\gamma T_i} \sum_{i=1}^N \int \frac{J_i^2}{P}dxdv.
\eeq
Note that $\Pi\ge 0$ (as expected). Conversely,
 the second term corresponds to the flux of entropy  given by
\beq
\Phi = -\sum_{i=1}^N \frac{m\kB}{T_i}\int v_i J_idxdv ,
\label{140}
\eeq
or even rewritten as
\beq
\Phi = \kB\sum_{i=1}^N \frac{\Phi^{(i)}_{\rm q}}{T_i}.
\label{14}
\eeq
As mentioned previously, Eq. (\ref{14}) can be alternatively used for evaluated
the steady production of entropy, since it depends only on averages
$\langle v_i^2\rangle $ and on  the temperatures $T_i$.


\section{Exact solution for time dependent  external forces} \label{sec3}
For simplifying matters, from now on we shall adopt $\kB=1$.
Except in Sec. \ref{tosc}, all  analysis will restrict to
the case of a chain of $N$ particles
interacting to its nearest neighbors
subject to an external force. The expression
for the force of $i$-th particle $f^{*}_{i}$ then reads
\begin{equation}
  f^{*}_{i} =  - \frac{k}{m} \left( x_{i} - x_{i+1} \right) - \frac{k^{'}}{m} x_{i} + f_{i}^{\text{ext}}(t) ,
\end{equation}
\begin{equation}
  f^{*}_{i} =  - \frac{k}{m} \left( x_{i} - x_{i-1} \right) - \frac{k^{'}}{m} x_{i} + f_{i}^{\text{ext}}(t) ,
\end{equation}
for  particles placed at extremities, $i=1$ and $N$, respectively, and
\begin{equation}
  f^{*}_{i} =  - \frac{k}{m} \left( 2x_{i} - x_{i-1}  - x_{i+1} \right) - \frac{k^{'}}{m} x_{i} + f_{i}^{\text{ext}}(t),
\end{equation}
for the intermediate ones.
Above expressions can be conveniently rewritten as
\begin{equation}
  f^{*}_{i} =  - K x_{i} + L x_{i+1} + f_{i}^{\text{ext}}(t) ,
\end{equation}
\begin{equation}
  f^{*}_{i} =  - K x_{i} + L x_{i-1} + f_{i}^{\text{ext}}(t) ,
\end{equation}
and, for $i=1$ and $i=N$, respectively, and
\begin{equation}
  f^{*}_{i} =  - \left( K + L \right)  x_{i} + L \left( x_{i+1} +  x_{i-1} \right)  + f_{i}^{\text{ext}}(t),
\end{equation}
respectively, where  $L = k/m$ and $K = (k + k^{'} )/m$ for $i\neq 1$ and $i\neq N$.

The time evolution of a generic average of type
$\langle g\rangle=\int g(x,y)P(x,v,t)dxdv$ is obtained
 through expression
\beq
\frac{d}{dt}\langle g\rangle=\int g(x,y) \frac{\partial P}{\partial t}dxdv,
\label{example1_2}
\eeq
and  by inserting  Eq. (\ref{3}) in Eq. (\ref{example1_2})
and performing appropriate partial integrations,
an explicit equation for the time evolution of
$\langle g\rangle$ is evaluated in terms of correlations associated
to the positions and velocities.
Due to the time dependence on the external forces,
the evaluation of averages like $\langle g\rangle$
 becomes
cumbersome. However, the calculations
  become quite simpler
 by rewriting  the motion
 equations in terms of  their associate covariances. For instance,
 let us take
 for example a generic average $\langle g \rangle=\expval{x_{i}^l x_{j}^m}$
 (with $l\ge 1$ and $m\ge 1$)
 with covariance given by
$\expval{x_{i}^l x_{j}^m}_{T}\equiv \langle {x_{i}^l x_{j}^m}\rangle -\langle x_{i}^l\rangle\langle x_{j}^m\rangle$. Unlike the time evolution of $\langle {x_{i}^l x_{j}^m}\rangle$,
the time equation for $d\expval{x_{i}^l x_{j}^m}_{T}/dt= d\langle {x_{i}^l x_{j}^m}\rangle/dt -\langle x_{j}^m\rangle d\langle x_{i}^l\rangle/dt-\langle x_{i}^l\rangle d\langle x_{j}^m\rangle/dt$
does not  depend explicitly on $t$.
Since the equations for all covariances are  linear and time independent,
the exact solution is possible for all system sizes $N$. Finally,
having the averages
$\expval{ v_{i}^{2} }_{T}$ and $\langle v_i\rangle^2$, the entropy flux
 can be directly evaluated
from the usage of Eqs. (\ref{11}) and (\ref{14}).

Below we derive  explicit expressions  for distinct covariances
between  the $i-$ and $i+1$-th particles for a generic chain of $N$ sites.
\begin{equation}
  \frac{d}{dt} \expval{ x_{i}^{2} }_{T} = 2\expval{v_{i} x_{i}}_{T},
  \label{qw1}
\end{equation}
\begin{equation}
      \frac{d}{dt} \expval{  x_{i} x_{i+1}}_{T} =  \expval{v_{i} x_{i+1} }_{T} + \expval{v_{i} x_{i+1} }_{T},
\end{equation}
\begin{equation}
      \frac{d}{dt} \expval{ x_{i} v_{i} }_{T} = \expval{ v_{i}^{2} }_{T} - K \expval{ x_{i}^{2} }_{T} + L \expval{ x_{i} x_{i+1} }_{T} - \gamma \expval{ x_{i}  v_{i}}_{T},
\end{equation}
\begin{equation}
  \begin{split}
    \frac{d}{dt} \expval{ x_{i} v_{i+1} }_{T} &= \expval{ v_{i} v_{i+1}}_{T} - K \expval{ x_{i} x_{i+1}  }_{T} + L \expval{ x_{i}^{2} }_{T} \\&- \gamma \expval{ x_{i}  v_{i+1}}_{T},
    \end{split}
\end{equation}
\begin{equation}
      \frac{d}{dt} \expval{ v_{i}^{2} }_{T} = - 2 K \expval{ v_{i} x_{i} }_{T} + 2 L  \expval{ x_{i+1} v_{i}  }_{T} -  2 \gamma \expval{  v_{i}^{2} }_{T} + \Gamma_{i},
\end{equation}
\begin{equation}
  \begin{split}
  \frac{d}{dt} \expval{ v_{i} v_{i+1} }_{T} &= -  K \expval{ x_{i} v_{i+1} }_{T} +  L  \expval{ x_{i+1} v_{i+1}  }_{T} \\
  & -  K \expval{ x_{i+1} v_{i} }_{T}+  L  \expval{ x_{i} v_{i}  }_{T} -  2 \gamma \expval{  v_{i} v_{i+1} }_{T}.
  \label{qw2}
  \end{split}
\end{equation}
Here we introduced the rescaled temperature $\Gamma_{i}$ defined by $\Gamma_{i} = 2 \gamma T_{i}/m$
and thereby for fixed
$\Gamma_{i}$'s, the achievement of $\expval{v_{i}^{2} }_{T}$'s
reduces  to  systems of linear equations.

The time evolution of single averages
$\langle v_i \rangle$ and $\langle x_i \rangle$ are also required for obtaining
$\langle v_i^2 \rangle$, whose expressions read
\begin{equation}
  \frac{d}{dt}\langle v_i\rangle=-(K+L)\langle x_{i}\rangle+L(\langle x_{i+1}\rangle
  +\langle x_{i-1}\rangle)-\gamma\langle v_{i}\rangle+f_{i}^{ext}(t),
  \label{v1}
\end{equation}
for $i\neq 1,N$ and
\begin{equation}
  \frac{d}{dt}\langle v_i\rangle=-K\langle x_{i}\rangle+L\langle x_{i+1}\rangle
  -\gamma\langle v_{i}\rangle+f_{i}^{ext}(t),
  \label{v11}
\end{equation}
\begin{equation}
  \frac{d}{dt}\langle v_i\rangle=-K\langle x_{i}\rangle+L\langle x_{i-1}\rangle
  -\gamma\langle v_{i}\rangle+f_{i}^{ext}(t),
  \label{v12}
\end{equation}
for $i=1$ and $N$, respectively and from Eq. (\ref{two_baths_mov2}),  the time
evolution of $\langle x_i\rangle$ reads
\begin{equation}
  \frac{d}{dt}\langle x_i\rangle=\langle v_i\rangle.
  \label{v2}
\end{equation}
Although the previous procedure does not depend
on the shape of external forces, from now on we will restrict our analysis  to harmonic
external forces given by $f_{i}^{ext}(t)=f_{0i}\cos(\omega t +\phi)$,
with  $\omega$ and $\phi$ being its frequency and phase difference (lag),
respectively.
By assuming that each $\expval{ x_{i} }$ has solution of type
$\expval{ x_{i} } = A_{0i} + A_{1i}  \cos \omega t + A_{2i} \sin \omega t$,
$\langle v_i\rangle$ becomes  $\langle v_i\rangle=\omega\Big[A_{2i}\cos(\omega t)-A_{1i}\sin(\omega t)\Big]$. By
inserting above solutions  in Eqs. (\ref{v1}) [or Eq. (\ref{v11})/(\ref{v12})] and (\ref{v2}), the coefficients $A_{1i}$ and $A_{2i}$
are obtained.
\section{Two particles case}
\subsection{Oscillating forces}
In the first application we analyze a chain of  two interacting
particles subjected to   harmonic forces without phase difference (lagless
case, $\phi=0$).
From the solution of linear set of equations described
in Sec. \ref{sec3} , we obtain the following
expressions for the covariances:
\begin{equation}
   \expval{ v_{1}^{2} }_{T} = \frac{\Gamma_{1} + \Gamma_{2} }{4 \gamma} + \frac{K \gamma \left( \Gamma_{1} - \Gamma_{2} \right)}{4 \left( L^{2} + K \gamma^{2} \right)},
\end{equation}
and
\begin{equation}
   \expval{ v_{2}^{2} }_{T} = \frac{\Gamma_{1} + \Gamma_{2} }{4 \gamma} - \frac{K \gamma \left( \Gamma_{1} - \Gamma_{2} \right)}{4 \left( L^{2} + K \gamma^{2} \right)}.
\end{equation}

The entropy flux $\Phi(t)$  can be splitted in two parts,
\begin{equation}
  \Phi(t) = \Phi_{T} + \Phi_{f}(t),
\end{equation}
where $\Pi_T=\Phi_{T}$ (in the steady state regime) and  $\Phi_{f}(t)$ read
\begin{equation}
  \Phi_{T} = \frac{2 \gamma^{2}}{\Gamma_{1}}  \expval{ v_{1}^{2} }_{T} + \frac{2 \gamma^{2}}{\Gamma_{2}}  \expval{ v_{2}^{2} }_{T}  - 2 \gamma,
  \label{expressao_1_2p}
\end{equation}
and
\begin{equation}
  \Phi_{f}(t) = \frac{2 \gamma^{2}}{\Gamma_{1}} \expval{v_{1}}^{2}+ \frac{2 \gamma^{2}}{\Gamma_{2}}  \expval{v_{2}}^{2},
\end{equation}
respectively.
Above expressions can be simplified, acquiring the following form
\begin{equation}
   \Pi_{T}=  \frac{ \gamma L^{2} }{ 2\left( L^{2} + K \gamma^{2} \right) } \frac{ \left( \Gamma_{1} - \Gamma_{1} \right)^{2} }{ \Gamma_{1} \Gamma_{2} },
\label{phit}
\end{equation}
and
\begin{equation}
  \Phi_{f} (t)=
  2 \gamma^2 \omega^2 \sum_{i} \left(\frac{(A_{2i} \cos \omega t-A_{1i} \sin \omega t)^2}{\Gamma_{i}} \right)
  \label{phif}
\end{equation}
respectively, whose coefficients $A_{1i}$ and $A_{2i}$ are shown in Appendix
\ref{apa}.
Note that $\Pi_{T}$ depends solely on  the difference of
temperatures and are similar to the case
with no external forces \cite{tome2010},
whereas $\Phi_{f}(t)$ is related to the time dependent forces.
By evaluating $\Phi_{f}(t)$ over a period cycle, we arrive at
the following expression for
${\overline \Pi}=\frac{\omega}{2\pi}\int_{0}^{2\pi/\omega}\Phi_{f}(t) dt$,
\begin{equation}
  \overline{\Pi} = \frac{\gamma^2 \omega^2 \left[\Gamma_{1} \left(A_{12}^2+A_{22}^2\right) +\Gamma_{2} \left(A_{11}^2+A_{21}^2\right) \right]}{\Gamma_{1} \Gamma_{2}}.
  \label{eq42}
  \end{equation}
\begin{widetext}
By substituting the expressions for $A_{1i}$'s and $A_{2i}$'s we finally
  arrive at the following expression:
\begin{equation}
  {\overline \Pi} = \frac{\gamma^2 \omega^2 \left\{f_{01}^2 \left[\Gamma_{2} \left(\gamma^2 \omega^2+\left(K-\omega^2\right)^2\right)+L^2 \Gamma_{1}\right]+2 f_{01} f_{02} L \left(K-\omega^2\right)
   (\Gamma_{1}+\Gamma_{2})+f_{02}^2 \left[\Gamma_{1} \left(\gamma^2 \omega^2+\left(K-\omega^2\right)^2\right)+L^2 \Gamma_{2}\right]\right\}}{\Gamma_{1} \Gamma_{2} \left[\gamma^2
      \omega^2+\left(K+L-\omega^2\right)^2\right] \left[\gamma^2 \omega^2+\left(-K+L+\omega^2\right)^2\right]}.
  \label{phif2}
\end{equation}
\end{widetext}
This is the one of  main results of the paper, and we pause to make
some few comments: First,
from Eq. (\ref{eq42}) it follows
that ${\overline \Pi}$ is always
greater than 0,
vanishing when $f_{01}=f_{02}=0$ and/or $\omega=0$. Second,
in the limit of slow or fast oscillations, $\omega<<1$ or $\omega>>1$,
${\overline \Pi}$ behaves as
\begin{equation}
  \overline{\Pi} \sim \frac{\gamma^2\Big[\Gamma_{1}(f_{01}K+f_{02}L)^2+\Gamma_{2}(f_{01}L+f_{02}K)^2\Big]\omega^2}{\Gamma_{1}\Gamma_{2}(K^2-L^2)^2},
\end{equation}
and
\begin{equation}
  {\overline \Pi}\sim \Big(\frac{f_{01}^2\Gamma_2+f_{02}^2\Gamma_1}{\Gamma_{1}\Gamma_{2}}\Big)\frac{\gamma^2}{\gamma^2+\omega^2},
\end{equation}
respectively, implying that ${\overline \Pi}$ vanishes as $\omega^2$
and $1/\omega^2$ for low and large periods, respectively. Third, when the
interaction between particles is ``weak'', $k<<k'$, ${\overline \Pi}$
reduces to the single  forced harmonic oscillator expression:
\begin{equation}
  \overline{\Pi} \sim\frac{\gamma^2\omega^2(f_{01}^2\Gamma_2+f_{02}^2\Gamma_1)}{\Gamma_1\Gamma_2[\gamma^2\omega^2+(K-\omega)^2]},
\end{equation}
acquiring  the simpler form \begin{equation}
  \overline{\Pi} \sim \frac{\gamma^2\omega^2f_{01}^2}{\Gamma_1[\gamma^2\omega^2+(K-\omega)^2]},
\end{equation}
\normalsize
as $\Gamma_1=\Gamma_2$ and $f_{01}=f_{02}$. Fourth and last,
in the strong coupling regime,
$k>>k'$ and $k/m>>\omega^2$ (or equivalently $L \approx K$ and $L>>\omega^2$),   $\overline{\Pi}$ becomes
\begin{equation}
  \overline{\Pi} \sim \Big(\frac{\gamma^2}{\gamma^2+\omega^2}\Big)\Big(\frac{\Gamma_1+\Gamma_2}{\Gamma_1\Gamma_2}\Big)(f_{01}+f_{02})^2,
  \label{aaa}
\end{equation}
which  is independent on
strength oscillator parameters $K$ and $L$.
\subsubsection{Bilinear form and Onsager coefficients}
The shapes of  Eqs. (\ref{phit}) and (\ref{phif2})
show that the entropy production components can be
written as flux-times-force expressions $\Pi_T={\cal J}_{T}f_T$
and
\begin{equation}
  {\overline \Pi}={\cal J}^{f}_1f_{01}+{\cal J}^{f}_2f_{02},
\end{equation}
respectively, where the forces     $f_T=1/\Gamma_1-1/\Gamma_2$ and $f_{0i(j)}$
have associated fluxes ${\cal J}_T$, ${\cal J}^{f}_1$ and  ${\cal J}^{f}_2$ given by
 \begin{equation}
    {\cal J}_T=\frac{\Gamma_1\Gamma_2 \gamma L^{2} }{ 2\left( L^{2} + K \gamma^{2}\right)}\Big(\frac{1}{\Gamma_1}-\frac{1}{\Gamma_2}\Big),
    \label{jt}
 \end{equation}
 and

\begin{equation}
{\cal J}^{f}_{1}=L_{11}f_{01}+L_{12}f_{02}, \quad {\rm and \quad} {\cal J}^{f}_{2}=L_{21}f_{01}+L_{22}f_{02},
\end{equation}
respectively. The  bilinear form for ${\overline \Pi}$ provides to identify
the terms $L_{11}$ and $L_{12}$ as the associated Onsager coefficients given by
\footnotesize{
  \begin{equation}
    L_{11}= \frac{\gamma^2\omega^2\left[\Gamma_{2} \left(\gamma^2 \omega^2+\left(K-\omega^2\right)^2\right)+L^2 \Gamma_{1}\right]}{\Gamma_1\Gamma_2\left[\gamma^2 \omega^2+\left(K+L-\omega^2\right)^2\right] \left[\gamma^2 \omega^2+\left(-K+L+\omega^2\right)^2\right]},
    \label{l11}
\end{equation}}
\normalsize
and
\footnotesize{
  \begin{equation}
    L_{12}=\frac{L \left(K-\omega^2\right)(\Gamma_{1}+\Gamma_2)}{\Gamma_1\Gamma_2\left[\gamma^2 \omega^2+\left(K+L-\omega^2\right)^2\right] \left[\gamma^2 \omega^2+\left(-K+L+\omega^2\right)^2\right]},
    \label{l12}
\end{equation}}
\normalsize
respectively. Analogous expressions are hold valid for $L_{21}$ and $L_{22}$
by exchanging $1\leftrightarrow 2$.
Note that $L_{11}\ge 0$ and  $L_{22}\ge 0$ (as expected).
The non-negativity of the entropy
 production also requires that $4L_{11}L_{22}-(L_{12}+L_{21})^2\ge 0$.
 To verify this, let us consider $\Gamma_2=r\Gamma_1$ with
 $r$ being an arbitrary (non negative) real number. Such above inequality is always satisfied, since the term
 \begin{equation*}
\frac{\left[\gamma^2 \omega^2+\left(K-\omega^2\right)^2+L^2 r\right] \left[r \left(\gamma^2 \omega^2+\left(K-\omega^2\right)^2\right)+L^2\right]}{L^2 (r+1)^2 \left(K-\omega^2\right)^2},
\end{equation*}
is greater than $1/4$ for all  values of $r,K,\omega$ and $\gamma$.
 \subsubsection{Phase difference between harmonic forces}
 Here we extend the results from previous section
 but taking into account a
 phase difference between external forces $f_1^{ext}(t)$ and $f_2^{ext}(t)$.
 More specifically, $f_1^{ext}(t)$  has the same expression as previously,
 but  $f_2^{ext}(t)$ now reads
 $f_2^{ext}(t)=f_{02}\cos(\omega t+\phi)$. By repeating
 aforedescribed procedures, we   assume that  $\expval{ v_{i} } = w  \left(  C_{2i}  \cos \omega t - C_{1i} \sin \omega t \right)$, whose  coefficients
 $C_{1i}$ and $C_{2i}$
 are decomposed in two parts:
 $C_{1i} = A_{1i} + B_{1i}(\phi)$, whose $A_{1i}$ and  $A_{2i}$ are the same as
 Eqs. (\ref{a1i}) and (\ref{a2i}) and the dependence on the   phase difference
 appears only in $B_{1i}$ and $B_{2i}$.   We then arrive at
the following expression for the steady entropy production  $\overline{\Pi}$
 \begin{equation}
   \overline{\Pi} = \frac{\gamma^2 \omega^2 \left[\Gamma_{1} \left(C_{12}^2+C_{22}^2\right) +\Gamma_{2} \left(C_{11}^2+C_{21}^2\right) \right]}{\Gamma_{1} \Gamma_{2}}.
 \end{equation}
 Above expression is quite similar to Eq. (\ref{eq42}) and
as in the lagless case, it has three terms  with
 first and third terms being identical to  Eq. (\ref{phif2}) and
 the phase difference dependence appearing only in middle term
 and given by

\begin{figure*}
  \centering
\includegraphics[scale=0.5]{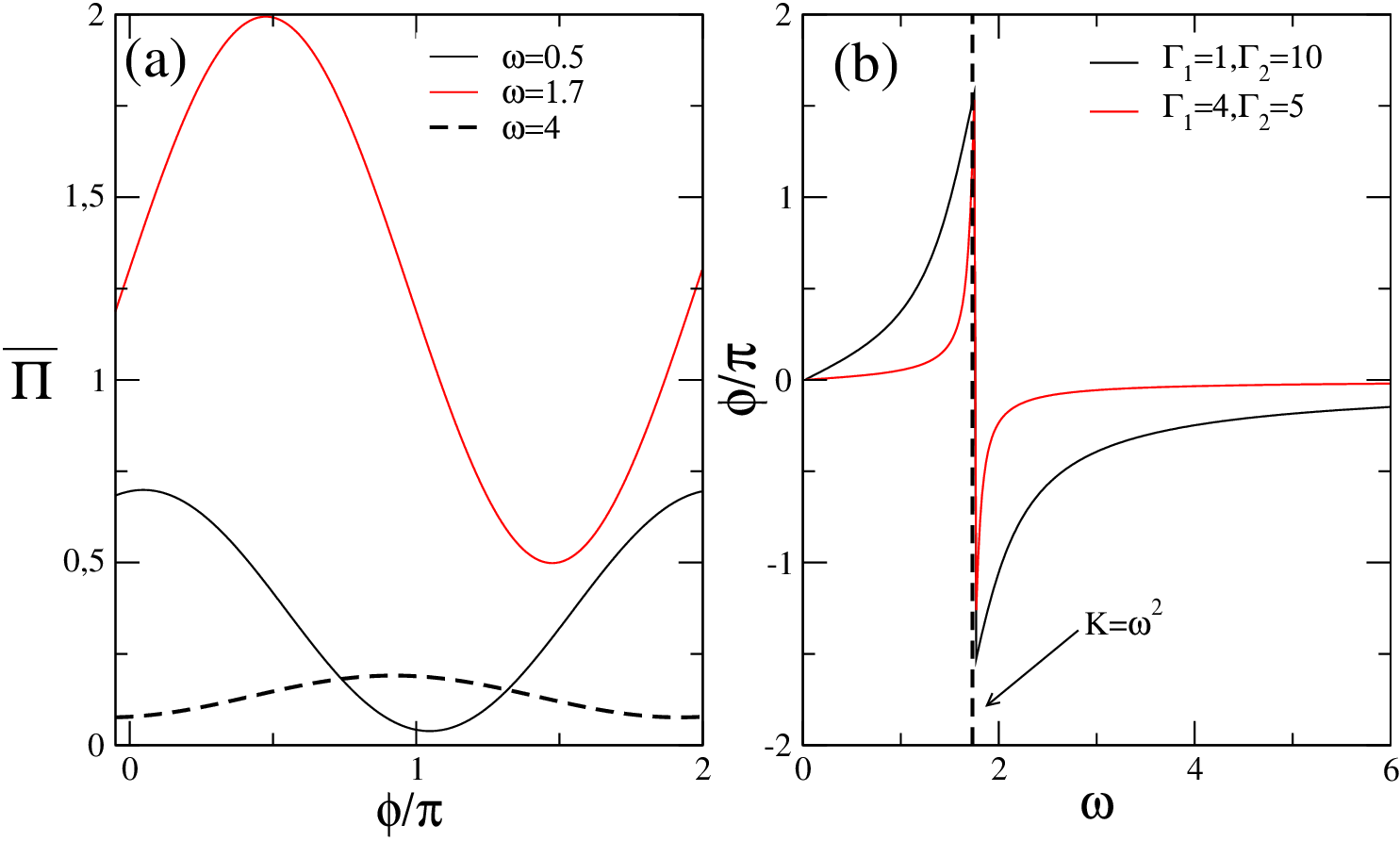}
\caption{ For distinct frequencies $\omega$'s, panel $(a)$ depicts the average
  entropy production ${\overline \Pi}$
  versus the phase difference $\phi$ for $\Gamma_{1}=1$, $\Gamma_{2}=10$,
  $K=3$ and $\gamma=1$.
  For  $K=3$ and $\gamma=1$  and distinct
  sets of $\Gamma_{1}$ and $\Gamma_{2}$, panel $(b)$ shows the positions $\phi$ of maximum/minimum of entropy production  ${\overline \Pi}$ versus $\omega$.}
\label{fig1}
\end{figure*}

 \footnotesize{
\begin{equation}
  \left[ \frac{ 2 f_{01} f_{02}  L  \left[\gamma \omega  (\Gamma_{2}-\Gamma_{1})\sin \phi+\left(K-\omega^2\right) (\Gamma_{1}+\Gamma_{2})\cos \phi \right]}{ \Gamma_{1} \Gamma_{2} \left(\gamma^2 \omega^2+\left(K+L-\omega^2\right)^2\right) \left(\gamma^2 \omega^2+\left(-K+L+\omega^2\right)^2\right)} \right].
 \end{equation}}
 \normalsize
 Note that the it reduces to the middle term from Eq. (\ref{phif2}) when $\phi=0$.
  The position of the maximum and minimum in the
  average entropy production   fulfills the above relation
\begin{equation}
  \phi= \tan^{-1} \left[  \frac{\gamma \omega (\Gamma_{2}-\Gamma_{1})}{\left(K-\omega^2\right) (\Gamma_{1}+\Gamma_{2})} \right].
  \label{extr}
\end{equation}
Note that $\phi$ depends only on the signs of both $\Gamma_{2}-\Gamma_{1}$ and $K-\omega^2$ and it is independent on $L$. In particular, in the regime of $\Gamma_{2}>>\Gamma_{1}(\Gamma_{2}<<\Gamma_{1})$, $\phi$ is independent on $\Gamma_i$'s, reading $\pm \gamma\omega/(K-\omega^2)$. Conversely,  for fast and slow
oscillations, it approaches to zero as $\gamma (\Gamma_{1}-\Gamma_{2})/K\omega(\Gamma_{2}+\Gamma_{1})$
and $\gamma\omega(\Gamma_{2}-\Gamma_{1})/K(\Gamma_{2}+\Gamma_{1})$, respectively.  Fig. \ref{fig1}  plots $\overline{\Pi}$ versus $\phi$ for distinct set
of values of $\omega$ and $\Gamma_{i}$'s. Note that the maxima
of mean entropy production yields at
$\phi \sim 0(\pi)$ for small (large) values of $\omega$
and $\phi \rightarrow \pi/2$ when $\omega \rightarrow {\sqrt K}$.
The dependence of extremes clearly follows theoretical
predictions from Eq. (\ref{extr}) (see e.g. panels $(a)$ and $(b)$ in Fig. \ref{fig1}).
\subsection{ Oscillating temperatures} \label{tosc}
The investigation of systems under oscillating temperature has been
reported in several works \cite{hohne2003,kraftmakher2004,filippov1966,sullivan1968,gobrecht1971,birge1985,gill1993,schawe1995,jeong1997,schawe1997,hohne1997,simon1997,jones1997,baur1998,claudy2000,garden2007a,garden2007c,garden2008}, which provides a way of measuring the heat
capacity experimentally. Here we intend to verify the nonequilibrium properties
of a chain of two interacting particles under time
oscillating temperatures.
For simplicity, we consider   external forces absent. The entropy production can also be evaluated straightforwardly from Eq. (\ref{14}), but instead the temperature
$\Gamma_i(t)$ of the each reservoir is now time dependent and given by
$\Gamma_i(t)=\Gamma_{0i}+F_{T_i}\cos(\omega t)$, with $\Gamma_{0i}$ and $F_{T_i}$ being the reference temperature and the strength of temperature driving, respectively.

\begin{figure*}
  \centering
\includegraphics[scale=0.5]{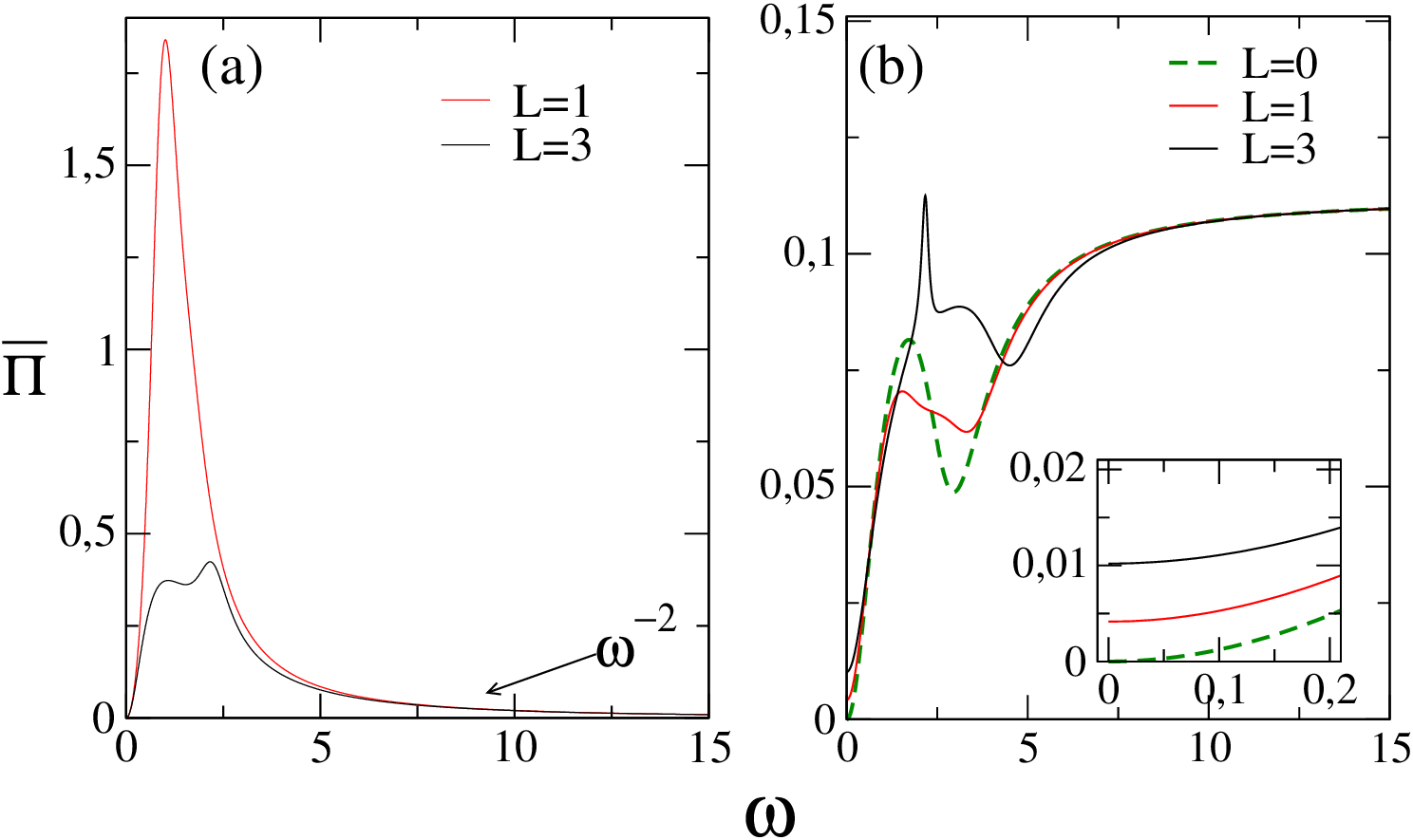}
\caption{ Panels $(a)$ and $(b)$ depict the steady ${\overline \Pi}$ versus
  frequency driving $\omega$ for time dependent oscillating forces and temperatures, respectively. In all cases we take $\gamma=1,K=2,\Gamma_1=\Gamma_2=10$ and
  $f_{02}=2f_{01}=4$ [panel $(a)$] and $F_{T_2}=2F_{T_1}=4$ [panel $(b)$]. Inset:
The steady ${\overline \Pi}$ for distinct $L$'s for low $\omega$.}
\label{fig12}
\end{figure*}

Although such problem is exactly solvable [see e.g. Fig. \ref{fig12} $(b)$] and reduces to the findings from Ref. \cite{fiore19} when $\Gamma_{01}=\Gamma_{02}$
and $L=0$, the expression for the steady
entropy production is  much more complex than previous cases and involves many terms related
to distinct powers of interaction parameters $K,L$ and driving
frequency $\omega$. For this reason, our analysis will be carried out close to equilibrium regime,
in which a linear treatment can be performed. More specifically, we take
both reference temperatures to be equal
$\Gamma_{01}=\Gamma_{02}=\Gamma_{0}$ and  the driving strengths are low $F_{T_i}<<\Gamma_{0}$. In such case, the entropy production ${\overline \Pi}$ can also be written down in the bilinear
form ${\overline \Pi}={\cal J}_{T_1}F_{T_1}+{\cal J}_{T_2}F_{T_2}$, where the fluxes ${\cal J}_{T_1}$
and ${\cal J}_{T_2}$ read
\begin{equation}
  {\cal J}_{T_1}=L_{T_1,T_1}F_{T_1}+L_{T_1,T_2}F_{T_2},
\end{equation}
and
\begin{equation}
   {\cal J}_{T_2}=L_{T_2,T_1}F_{T_1}+L_{T_2,T_2}F_{T_2},
\end{equation}
respectively, where $L_{T_i,T_j}$ are the associated Onsager coefficients given by
\begin{equation}
  L_{T_1,T_1}=\frac{\gamma}{2\Gamma_0^2}\left( \frac{ \sum_{\ell=0}^{9} B_{\ell} \ \omega^{2\ell}}{\sum_{\ell=0}^{9} G_{\ell} \ \omega^{2\ell}}\right),
\end{equation}
and
\begin{equation}
  L_{T_1,T_2}=\frac{-4\gamma^3L^2}{\Gamma_0^2}\left( \frac{ \sum_{\ell=0}^{6} A_{\ell} \ \omega^{2\ell}}{\sum_{\ell=0}^{9} G_{\ell} \ \omega^{2\ell}}\right),
\end{equation}
respectively, where $L_{T_1,T_1}= L_{T_2,T_2}$ and $L_{T_1,T_2}= L_{T_2,T_1}$ and coefficients $A_i$'s, $B_i$'s and $G_i$'s depend solely on
the parameters $\gamma$ and $L=2K$ and are listed in the Appendix \ref{apc}.

We pause again to make some comments: First, in the limit of slow and fast
frequencies,
${\overline \Pi}$ approaches to the following expressions
\begin{equation}
  {\overline \Pi} \sim \frac{4\gamma^3L^2A_0}{\Gamma_0^2G_0}(F_{T_1}-F_{T_2})^2,
   \label{eq1244}
\end{equation}
and
\begin{equation}
  {\overline \Pi} \sim \frac{\gamma}{2\Gamma_0^2}(F_{T_1}^2+F_{T_2}^2),
  \label{eq1245}
\end{equation}
respectively.
They  contrast with the  oscillating forced case, since are independent
on  $\omega$ and  different from zero in both extreme cases.
Whenever it depends on $L$ for low oscillations, the entropy production is independent on the coupling for fast oscillations.
Finally, for strong interaction strength, $L>>1$ and $L>>\omega^2$,  ${\overline \Pi}$
reads
\begin{equation}
  {\overline \Pi} \sim \frac{\gamma}{4\Gamma_0^2(\gamma^2+\omega^2)}
  \left[\gamma^2(F_{T_1}-F_{T_2})^2+2\omega^2(F_{T_1}^2+F_{T_2}^2)\right],
  \label{eq1246}
\end{equation}
which is also independent on $L$.
We close this section by comparing, in Fig. \ref{fig12} $(a)$ and $(b)$ the steady entropy production behaviors versus
the frequency driving $\omega$ for both oscillating
and temperature forces (obtained from the exact solution).
They exhibit meaningfully different dependence on $\omega$, even for extreme
$\omega$. Whenever ${\overline \Pi}$ vanishes for $\omega<<1$ and $\omega>>1$
in the case of time oscillating forces, it reaches  constant
values for temperature drivings,
in accordance with asymptotic expressions Eqs. (\ref{eq1244}) and (\ref{eq1245}), respectively, 
obtained from the linear regime approximation.
\section{More than two particles}
In this section we present the main results for long  chains
of oscillators, in
which the difference of temperature between particles
placed at the extremities is responsible for a transport  of heat
following Fourier's law. More concretely, it states that the heat current
is proportional to the inverse of the length of the chain given by
\begin{equation}
  {\cal J}_T=-\kappa\frac{dT}{dx},
\end{equation}
where $\kappa$ is the heat conductivity. In
the case of a finite difference of temperatures $\Delta T$, it follows that ${\cal J}_T\sim 1/N$ and thereby the heat flux is proportional to the inverse
of the system size.
Obtaining Fourier's Law from
 microscopic models have attracted great interest of
in the last years \cite{lepri,dhar,landi1,mario2}.  One of
the simplest model   consists of  a chain of interacting atoms
in contact with heat reservoirs at different temperatures
at its ends.
Among the distinct approaches aimed at  obtaining a heat flux inversely
proportional to the system chain,  we mention the self-consistent
protocol
proposed by Bosterli et al. \cite{boster}. More specifically,
 the temperature of the $i$-th intermediate
reservoir is chosen so that  it does not exchange heat
with the system.
Here we take a somewhat different approach by Bosterli et al.
in which each intermediate temperature is chosen so that the
variance $\expval{ v_{i}^{2} }_{T}$ is given by
$\expval{ v_{i}^{2} }_{T}= \Gamma_{i}/(2 \gamma)$.
In contrast to Ref. \cite{boster},
this self-consistent choice  leads to dissipation of heat
due to the external forces and consequently
all self-consistent reservoirs are expected to  produce entropy.

In order to compare the distinct sources of dissipation,
  thermal and time oscillating forces, we will
 consider that particles placed at extremities
 are not subjected to external forces.
 Thereby, under the above choice the intermediate $\Gamma_i$'s,
 the flux of entropy becomes
 \begin{equation}
  \Phi(t) = \Phi_{T} + \Phi_{f}(t),
\end{equation}
where $\Phi_{T}$  read
\begin{equation}
  \Phi_{T} = \frac{2 \gamma^{2}}{\Gamma_{1}}  \expval{ v_{1}^{2} }_{T} + \frac{2 \gamma^{2}}{\Gamma_{N}}  \expval{ v_{N}^{2} }_{T}  - 2 \gamma,
\end{equation}
and  $\Phi_{f}(t)$ is a sum of individual contributions
\begin{equation}
  \Phi_{f}(t)= 2 \gamma^{2}\sum_{i=1}^N\frac{\expval{v_{i}}^{2}}{\Gamma_{i}}.
  \label{longn}
\end{equation}
It is worth mentioning that despite the absence of external forces for extreme particles, the averages
$\langle v_{1}\rangle$ and $\langle v_{N}\rangle$
present oscillating behavior due to the couplings with neighboring particles (see e.g. Eqs. (\ref{v11}) and (\ref{v12})).

Finally, in all cases the steady entropy production rate $\Pi_T=\Phi_{T}$ can be written in the following form
\begin{equation}
  \Pi_{T}={\cal J}_Tf_T,
  \label{phitn}
\end{equation}
where the thermodynamic force $f_T$ and its associate flux ${\cal J}_T$ read $f_T=1/\Gamma_N-1/\Gamma_1$ and $ {\cal J}_T=-\kappa(\Gamma_N-\Gamma_1)/N$, respectively.
Thereby, the expression for $\Pi_{T}$ becomes
\begin{equation}
  \Pi_{T}=\frac{\kappa}{N}\frac{(\Gamma_N-\Gamma_1)^2}{\Gamma_1\Gamma_N}.
  \label{kappa}
\end{equation}
Since    the thermal conduction coefficient $\kappa$
is finite (it depends only on parameters $\Gamma_1,\Gamma_N,K$ and $L$),
the entropy production $\Pi_T$ decays as $N^{-1}$ (see e.g. Fig. \ref{fig3}$(b)$).

\subsection{Three particles}
Here we derive explicit results for a chain of $N=3$ particles. In such
case, Eq. (\ref{longn}) becomes
\begin{equation}
  \Phi_{f}(t) = \frac{2 \gamma^{2}}{\Gamma_{1}} \expval{v_{1}}^{2}+ \frac{2 \gamma^{2}}{\Gamma_{2}}  \expval{v_{2}}^{2} + \frac{2 \gamma^{2}}{\Gamma_{3}}  \expval{v_{3}}^{2},
\end{equation}
and the entropy production $\Pi_{T}$ due to thermal reservoirs
has the shape from Eq. (\ref{phitn}) with $ {\cal J}_T$
given by 
\begin{equation}
 {\cal J}_T  =  \frac{ \Gamma_1\Gamma_3\gamma L^{2} \left( 2 \gamma^{2} K + L^{2} \right) }{ 2 \left[ L^{2} + \gamma^{2} \left( 4 K - 2L \right)  \right] \left[ L^{2} + \gamma^{2} \left( K + L \right) \right] }f_T.
\end{equation}
Once again, $\Pi_{T}\ge 0$, since
$ 4 K - 2L=2(k+2k')/m$.
Using the motion equations we arrive at the following
expression for ${\overline \Pi}$:
\begin{widetext}
\begin{equation}
  \overline{\Pi} = \frac{f_{02}^2 \gamma^2 \omega^2 \left[\Gamma_{1} \Gamma_{3} \left(\gamma^2 \omega^2+\left(K-\omega^2\right)^2\right)+L^2 \Gamma_{2} (\Gamma_{1}+\Gamma_{3})\right]}{\Gamma_{1} \Gamma_{2} \Gamma_{3} \left[\gamma^2
   \omega^2+\left(K+2 L-\omega^2\right)^2\right] \left[\gamma^2 \omega^2+\left(-K+L+\omega^2\right)^2\right]},
\end{equation}
\end{widetext}
which is strictly positive and vanishes when $f_{02}$ and/or $\omega$
are equal to zero. Also, in the regime of slow and fast oscillations,
$\overline{\Pi}$ exhibit similar dependences on $\omega$ to
the two particles case:
\begin{equation}
  \overline{\Pi} \sim \Big(\frac{K^2}{\Gamma_2}+\frac{L^2(\Gamma_1+\Gamma_3)}{\Gamma_1\Gamma_3}\Big)\frac{\gamma^2\omega^2f_{02}^2}{(K+2L)^2(-K+L)^2},
\end{equation}
for $\omega<<1$ and
\begin{equation}
  \overline{\Pi}\sim \frac{\gamma^2f_{02}^2}{\Gamma_2}\frac{1}{\omega^2},
\end{equation}
for $\omega>>1$, respectively, implying that for such latter
limit the entropy production
is independent on extreme temperatures.
For strong couplings between
particles, $L \approx K>>\omega^2$,
$\overline{\Pi}$
approaches to
\begin{equation}
  \overline{\Pi}\sim \frac{f_{02}^2\gamma^2}{\gamma^2+\omega^2}\frac{(\Gamma_1\Gamma_2+\Gamma_1\Gamma_3+\Gamma_2\Gamma_3)}{\Gamma_1\Gamma_2\Gamma_3},
\end{equation}
which, in similarity with Eq. (\ref{aaa}) [for $N=2$], $\overline{\Pi}$
is independent on the interaction strengths.
\subsection{The limit of long particle chains}
\begin{figure}
  \centering
\includegraphics[scale=0.3]{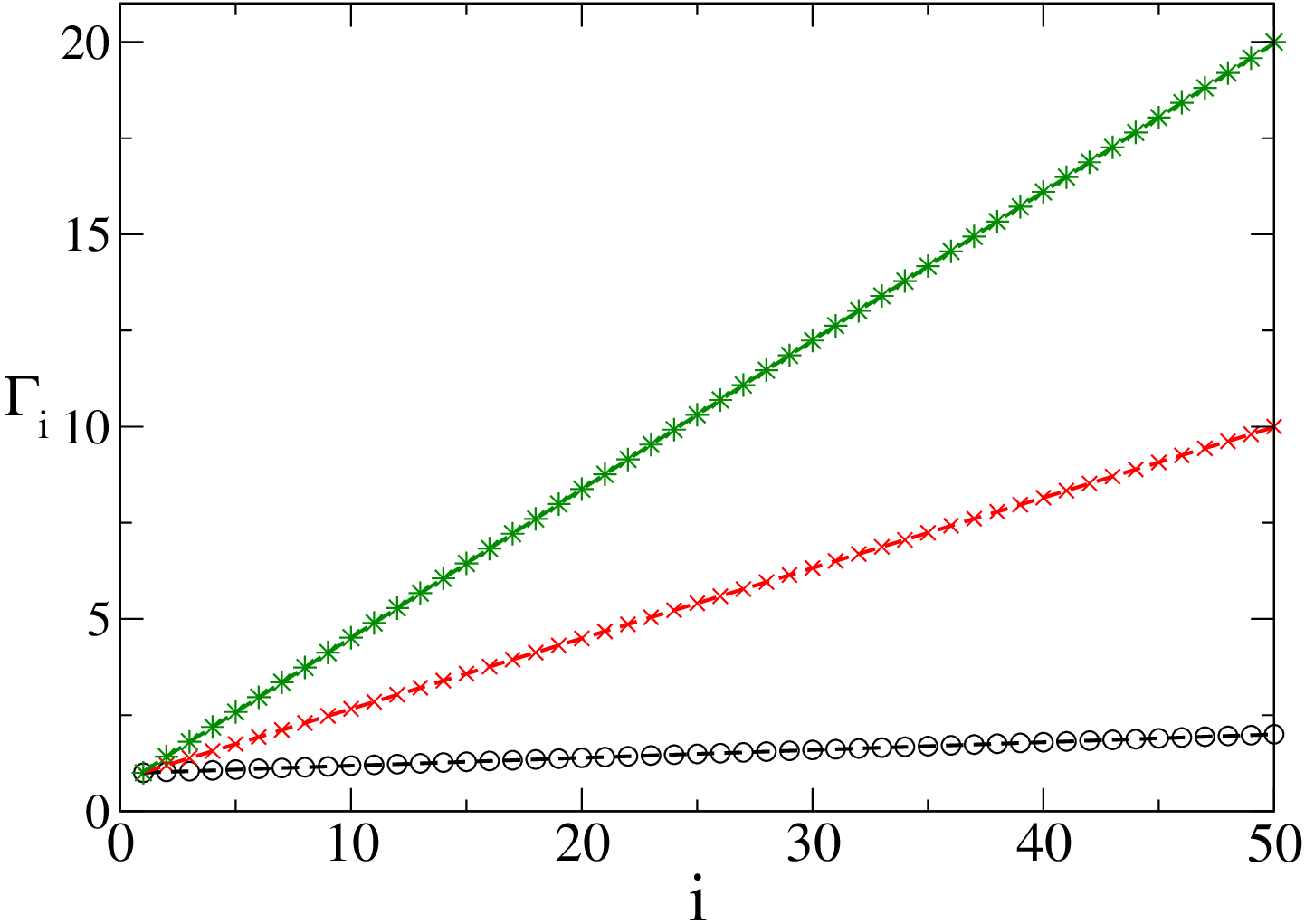}
\caption{ For a chain of $N=50$ particles with $K=2,L=1,\gamma=1$
  and $\omega=1$, the rescaled temperatures  $\Gamma_i$'s versus the position of the $i$-th site
  for three set of  temperatures $(\Gamma_1,\Gamma_N)$. The
intermediate temperatures are calculated according to the prescription $\expval{ v_{i}^{2} }_{T}= \Gamma_{i}/(2 \gamma)$.}
\label{fig2}
\end{figure}

All results obtained for $N=3$ particles
can be straightforward extended for long chains.
However, it becomes very cumbersome  to obtain simplified
expressions for ${\overline \Pi}$ in such cases. For this reason, we will
restrict the next analysis for specific values of control parameters.
Fig. \ref{fig2} shows,  for a chain of $N=50$ particles and
three sets of temperatures ($\Gamma_1,\Gamma_N$),
the temperature profiles calculated from the self consistent
protocol. In all cases,
the set of intermediate temperatures changes linearly from $\Gamma_1$ to $\Gamma_N$, consistent to a flux of heat along the chain from the hot to the cold reservoirs.

\begin{figure*}
  \centering
\includegraphics[scale=0.5]{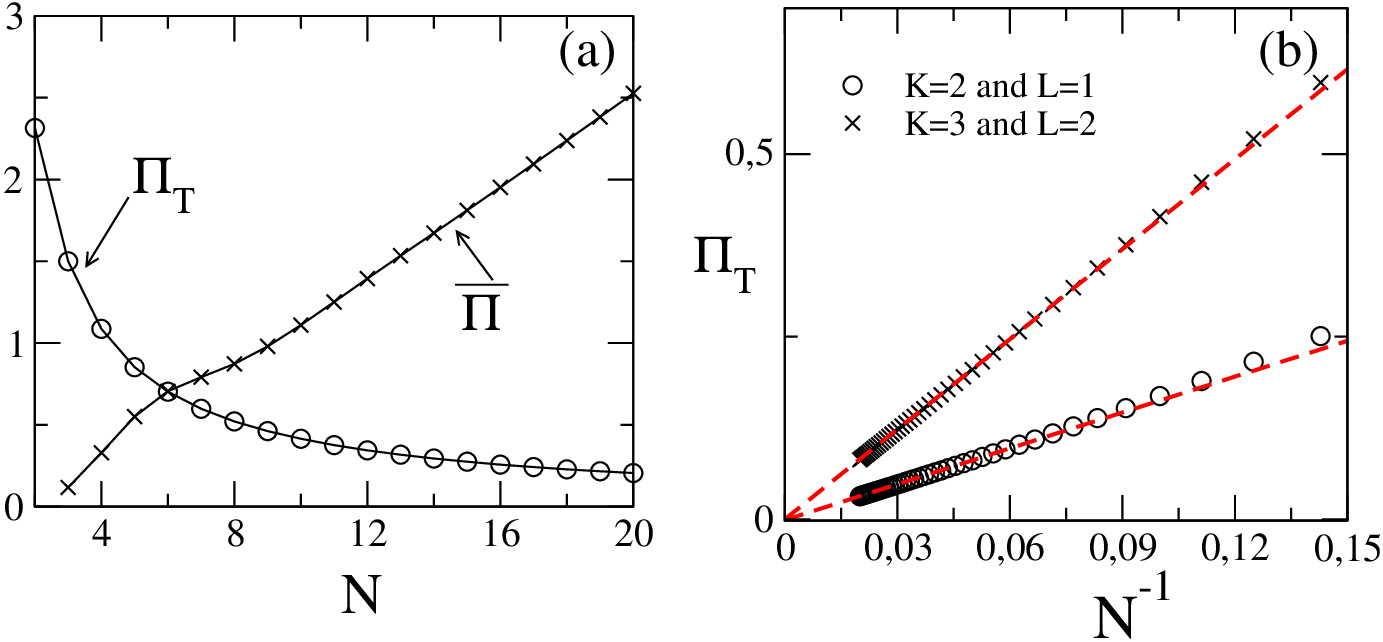}
\caption{ For $\Gamma_1=1$ and $\Gamma_N=10$, panels $(a)$  depicts
  the individual entropy
  production contributions from the thermal and
  self-consistent baths
  versus  $N$ for $K=3,L=2$ and $\omega=1.5$. In $(b)$ the
behavior of  entropy production from the thermal reservoirs $\Pi_T$ vs $N^{-1}$.}
\label{fig3}
\end{figure*}
Fig. \ref{fig3}$(a)$ compares the individual entropy
production contributions for distinct system sizes for
$f_{02}=f_{03}=...=f_{0N-1}$. Since  external
forces are equally presented in all intermediate  particles,  the
entropy production
  associated to self consistent
  baths increases
  linearly with $N$. Also, panel $(a)$ depicts the existence
  of two regimes. For small chains the thermal reservoir contribution $\Pi_T$ dominates over the self-consistent ones
  ${\overline \Pi}$, whereas
  ${\overline \Pi}$ wins over $\Pi_T$ upon $N$ is increased. In the limit
  $N \rightarrow \infty $ (see e.g panel $(b)$), only the contributions from
  self-consistent reservoirs prevail, in consonance with Fourier's
  law, Eq. (\ref{kappa}). Finally, it is worth emphasizing two
  distinct linear behaviors of ${\overline \Pi}$.
  It arises from  the particles closer
  to the thermal
  reservoirs   providing more contribution for the entropy
  production for small chains than for large ones.


  \section{Conclusions}
  The nonequilibrium properties of linear chains of Brownian particles
  were analyzed via stochastic thermodynamics.
  Expressions for the heat flux, entropy production and allied quantities
  were exactly obtained. The regimes of short and long chains
  were detailed inspected. In the former case the entropy production
  was derived as  bilinear functions of fluxes and forces,
  from which the associated Onsager coefficients depend on
  interaction couplings and frequency drivings.
  Reciprocal relations were also obtained. The limit of long
  chains  was studied by means of a self-consistent
  protocol for choosing intermediate temperatures. The entropy
  production is a sum of two terms: one coming from the real
  baths and the other from the self-consistent reservoirs.
  Whenever the former dominates for short chains, the latter
  contribution prevails for long ones. The contribution from the
  thermal reservoirs is responsible to heat flux according to Fourier's law.

  As a final comment, it is worth to discuss future extensions of the
  present study. The inclusion of both temperature and external force
  drivings in harmonic chains should be interesting, in order to compare
  not only the structure of entropy production  but also the Onsager
  coefficients. Also, the investigation of other kinds of drivings,
  such as the time discrete drivings should also be interesting
  in order to compare with sinusoidally time dependent ones.




  \section{Acknowledgment}
  We acknowledge Karel Proesmans for a careful reading of the manuscript.
C. E. F acknowledges the financial support from FAPESP under grant
  2018/02405-1.

  \appendix

  \section{Obtaining the coefficients $A_{ij}$ for $N=2$ particles case with no phase difference}\label{apa}
  Here we show explicit expressions for the coefficients $A_{ij}$'s for the
  two particles case subject to
oscillating forces. The index $i$ stands for the $i$-th particle ($i=1,2$).

  \begin{widetext}
  \begin{equation}
    A_{1i} = \frac{f_{0i} \left(K-\omega^2\right) \left[\gamma^2 \omega^2+\left(K-\omega^2\right)^2-L^2\right]+f_{0j} L \left[-\omega^2 \left(\gamma^2+2 K\right)+K^2-L^2+\omega^4\right]}{\left[\gamma^2 \omega^2+\left(K+L-\omega^2\right)^2\right] \left[\gamma^2 \omega^2+\left(-K+L+\omega^2\right)^2\right]},
    \label{a1i}
 \end{equation}
\end{widetext}
and
\footnotesize{
  \begin{equation}
    A_{2i} = \frac{\gamma \omega \left[f_{0i} \left(\gamma^2 \omega^2+\left(K-\omega^2\right)^2+L^2\right)+2 f_{0j} L \left(K-\omega^2\right)\right]}{\left(\gamma^2 \omega^2+\left(K+L-\omega^2\right)^2\right) \left(\gamma^2 \omega^2+\left(-K+L+\omega^2\right)^2\right)}.
      \label{a2i}
\end{equation}}
\normalsize

Having the $A_{ij}$'s, the steady entropy production ${\overline \Pi}$ is straightforwardly evaluated.
  \section{Obtaining the coefficients $B_{ij}$ for $N=2$ particles case and phase difference}\label{apb}
  Here we show explicit expressions for the coefficients $B_{ij}$'s
  for the two particles subject to
  phase difference between oscillating forces.
  \begin{widetext}
     \begin{equation}
      B_{11} = \frac{f_{02} L \left((\cos \phi-1) \left(-\omega^2 \left(\gamma^2+2 K\right)+K^2-L^2+\omega^4\right)+2 \gamma \omega \left(K-\omega^2\right) \sin \phi\right)}{\left(\gamma^2 \omega^2+\left(K+L-\omega^2\right)^2\right) \left(\gamma^2 \omega^2+\left(-K+L+\omega^2\right)^2\right)},
  \end{equation}
\end{widetext}

  \begin{widetext}
     \begin{equation}
      B_{21} = \frac{f_{02} L \left(\sin \phi \left(\gamma^2 \omega^2-\left(K-\omega^2\right)^2+L^2\right)+2 \gamma \omega \left(K-\omega^2\right) (\cos \phi-1)\right)}{\left(\gamma^2 \omega^2+\left(K+L-\omega^2\right)^2\right) \left(\gamma^2 \omega^2+\left(-K+L+\omega^2\right)^2\right)},
  \end{equation}
\end{widetext}

  \begin{widetext}
  \begin{equation}
B_{12} = \frac{f_{02} \gamma \omega \sin \phi \left(\gamma^2 \omega^2+\left(K-\omega^2\right)^2+L^2\right)+f_{02} \left(K-\omega^2\right) (\cos \phi-1) \left(\gamma^2 \omega^2+\left(K-\omega^2\right)^2-L^2\right)}{\left(\gamma^2 \omega^2+\left(K+L-\omega^2\right)^2\right) \left(\gamma^2 \omega^2+\left(-K+L+\omega^2\right)^2\right)},
  \end{equation}
  \end{widetext}
and
  \begin{widetext}
  \begin{equation}
B_{22} = \frac{f_{02} \left(\omega^2-K\right) \sin \phi \left(\gamma^2 \omega^2+\left(K-\omega^2\right)^2-L^2\right)+f_{02} \gamma \omega (\cos \phi-1) \left(\gamma^2 \omega^2+\left(K-\omega^2\right)^2+L^2\right)}{\left(\gamma^2 \omega^2+\left(K+L-\omega^2\right)^2\right) \left(\gamma^2 \omega^2+\left(-K+L+\omega^2\right)^2\right)},
  \end{equation}
 \end{widetext}
respectively.
  Note that all of them vanishes as $\phi=0$, restoring the expressions
  Eqs. (\ref{a1i}) and (\ref{a2i}), respectively.

  \section{Obtaining the coefficients $A_{i}$'s, $B_i$'s and $G_i$'s for $N=2$ particles case and
    oscillating temperature}\label{apc}
  Here we show explicit expressions for the coefficients $A_{i}$'s, $B_i$'s and $G_i$'s for the
  two particles case and time oscillating temperatures.
  \begin{widetext}

    \begin{flalign*}
    &
    A_{0} = 2304 L^5 \left(2 \gamma^2+L\right),
    &
    \end{flalign*}
    \begin{flalign*}
    &
    A_{1}  = -128 L^3 \left(-7 \gamma^4+58 \gamma^2 L+123 L^2\right),
    &
    \end{flalign*}
    \begin{flalign*}
    &
    A_{2}  = 16 L \left(8 \gamma^6-50 \gamma^4 L+34 \gamma^2 L^2+931 L^3\right),
    &
    \end{flalign*}
    \begin{flalign*}
    &
    A_{3}  = 4 \left(-11 \gamma^6+78 \gamma^4 L+319 \gamma^2 L^2-1606 L^3\right),
    &
    \end{flalign*}
    \begin{flalign*}
    &
    A_{4}  = -3 \left(17 \gamma^4+82 \gamma^2 L-548 L^2\right),
    &
    \end{flalign*}
    \begin{flalign*}
    &
    A_{5}  = 3 \gamma^2-214 L,
    &
    \end{flalign*}
    \begin{flalign*}
    &
    A_{6}  = +10,
    &
    \end{flalign*}
    \begin{flalign*}
    &
    B_{0} = 36864 \gamma^4 L^7+18432 \gamma^2 L^8,
    &
    \end{flalign*}
    \begin{flalign*}
    &
    B_{1} =  31744 \gamma^6 L^5+112640 \gamma^4 L^6+27648 \gamma^2 L^7+36864 L^8,
    &
    \end{flalign*}
    \begin{flalign*}
    &
    B_{2} = 4608 \gamma^8 L^3+18176 \gamma^6 L^4-77056 \gamma^4 L^5+203904 \gamma^2 L^6-172032 L^7,
    &
    \end{flalign*}
    \begin{flalign*}
    &
    B_{3} = 128 \gamma^{10} L+608 \gamma^8 L^2-6592 \gamma^6 L^3+85920 \gamma^4 L^4-230720 \gamma^2 L^5+269824 L^6,
    &
    \end{flalign*}
    \begin{flalign*}
    &
    B_{4} =  16 \gamma^{10}-64 \gamma^8 L+4536 \gamma^6 L^2-34864 \gamma^4 L^3+125488 \gamma^2 L^4-170496 L^5,
    &
    \end{flalign*}
    \begin{flalign*}
    &
    B_{5} = 56 \gamma^8-792 \gamma^6 L+8112 \gamma^4 L^2-34928 \gamma^2 L^3+54288 L^4,
    &
    \end{flalign*}
    \begin{flalign*}
    &
    B_{6} = 73 \gamma^6-928 \gamma^4 L+5120 \gamma^2 L^2-9536 L^3,
    &
    \end{flalign*}
    \begin{flalign*}
    &
    B_{7} = 43 \gamma^4-376 \gamma^2 L+936 L^2,
    &
    \end{flalign*}
    \begin{flalign*}
    &
    B_{8} = 11 \gamma^2-48 L,
    &
    \end{flalign*}
    \begin{flalign*}
    &
    B_{9} = 1,
    &
    \end{flalign*}
    \begin{flalign*}
    &
    G_{0} = 147456 \gamma^6 L^6+147456 \gamma^4 L^7+36864 \gamma^2 L^8,
    &
    \end{flalign*}
    \begin{flalign*}
    &
    G_{1} = 50176 \gamma^8 L^4-94208 \gamma^6 L^5+262144 \gamma^4 L^6-24576 \gamma^2 L^7+36864 L^8,
    &
    \end{flalign*}
    \begin{flalign*}
    &
    G_{2} = 3584 \gamma^{10} L^2-23552 \gamma^8 L^3+166400 \gamma^6 L^4-323584 \gamma^4 L^5+384512 \gamma^2 L^6-172032 L^7,
    &
    \end{flalign*}
    \begin{flalign*}
    &
    G_{3} = 64 \gamma^{12}-768 \gamma^{10} L+14720 \gamma^8 L^2-77312 \gamma^6 L^3+262528 \gamma^4 L^4-399872 \gamma^2 L^5+269824 L^6,
    &
    \end{flalign*}
    \begin{flalign*}
    &
    G_{4} = 240 \gamma^{10}-2688 \gamma^8 L+24672 \gamma^6 L^2-96960 \gamma^4 L^3+200592 \gamma^2 L^4-170496 L^5,
    &
    \end{flalign*}
    \begin{flalign*}
    &
    G_{5} = 348 \gamma^8-3504 \gamma^6 L+20024 \gamma^4 L^2-52736 \gamma^2 L^3+54288 L^4,
    &
    \end{flalign*}
    \begin{flalign*}
    &
    G_{6} = 245 \gamma^6-2064 \gamma^4 L+7424 \gamma^2 L^2-9536 L^3,
    &
    \end{flalign*}
    \begin{flalign*}
    &
    G_{7} = 87 \gamma^4-528 \gamma^2 L+936 L^2,
    &
    \end{flalign*}
    \begin{flalign*}
    &
    G_{8} = 15 \gamma^2-48 L,
    &
    \end{flalign*}
    \begin{flalign*}
    &
    G_{9} = 1.
    &
    \end{flalign*}

 \end{widetext}

\end{document}